# Experimental Confirmation of Quantum Hall Ferromagnetic State in an Organic Dirac Fermion System

Mitsuyuki Sato, Takako Konoike,[#] and Toshihito Osada[*]

*Institute for Solid State Physics, University of Tokyo, Kashiwa, Chiba 277-8581, Japan.*

We have experimentally confirmed the quantum Hall ferromagnetic state with Chern number $\nu = 0$, characterized by the helical edge state, in a layered organic Dirac fermion system $\alpha$-(BEDT-TTF)$_2$I$_3$. The interlayer resistance saturates at low temperatures and high magnetic fields. It does not scale with the sample cross-sectional area in the saturating region, and resonantly depends on the magnetic field direction. These results strongly suggest that the helical edge state dominates transport. This is the first observation of the topological phase in organic molecular crystals.

For the past fifteen years, the two dimensional (2D) massless Dirac fermion systems such as graphene have been one of the important subjects in condensed matter physics [1]. Generally, Dirac cones have twofold spin degeneracy and appear as a pair at the time-reversal-symmetric points in the Brillouin zone, called valleys. One of the most characteristic features of 2D massless Dirac fermions under the magnetic field is the ground Landau level (LL) labeled $n=0$, the energy of which is always equal to the Dirac point energy, namely zero. The $n=0$ LL has fourfold degeneracy with respect to spin and valley, and it splits to four levels due to interaction and the Zeeman effect under sufficiently high magnetic field [2]. The quantum Hall (QH) state with a Chern number of zero ($v=0$), in which the in-plane Hall conductivity is quantized to zero, appears when the Fermi level is located in the central mobility gap in the four levels.

Different types of the $v=0$ QH ground state can be considered depending on how to break the spin and valley degeneracy. When the spin splitting is dominant, the $v=0$ state is a spin-polarized state called the QH ferromagnet (QHF) state. On the other hand, when valley splitting is dominant, the $v=0$ state is the spin-unpolarized state accompanied by spin or charge density modulation, which is called the QH insulator (QHI) state [3]. One of the most remarkable differences between the QHF and QHI states is the existence of a helical edge channel. Figure 1(a) illustrates the energy dispersion of the split levels of the $n=0$ LL in the 2D QHF state. Here, we ignore the interaction for simplicity. The $n=0$ LL shows the splitting into two levels with spin $\sigma_z=+1$ and $\sigma_z=-1$ in the bulk region, and each spin level splits into two branches around the edge due to the edge potential. One of the two branches of each spin crosses the Fermi level, and forms a chiral edge channel surrounding the 2D system as illustrated in Fig. 1(b). The pair of chiral edge channels with opposite spin and group velocity is the helical edge channel [4,5]. It is topologically protected as long as the spin component along the magnetic field ($\sigma_z$) is conserved. On the other hand, there are no protected gapless edge states in the QHI state.

In monolayer graphene, the possible $\nu = 0$ QH states were first discussed by Kharitonov using a renormalization group approach. In addition to the QHF states, three types of the QHI state appear, depending on the anisotropic interaction energy and the Zeeman energy; the canted antiferromagnet (CAF) state, the Kekule distortion state, and the charge density wave state [6,7]. It has been experimentally confirmed that the CAF state is realized as the $\nu = 0$ QH state under normal magnetic fields. On the other hand, the QHF state appears under highly tilted magnetic fields where the Zeeman splitting becomes dominant [8]. In bilayer graphene, which has the ground LL with eightfold degeneracy, much richer symmetry-breaking states have been discussed [9-12].

In this paper, we report the experimental confirmation of surface transport in the $\nu = 0$ QH states in the 2D Dirac fermion system in a layered organic conductor $\alpha$-(BEDT-TTF)$_2$I$_3$, where BEDT-TTF denotes bis(ethylenedithio)-tetrathiafulvalene. This result indicates that the QHF state with helical edge state is stable under the normal magnetic field, in contrast to graphene. We investigated the scaling and angle-dependent features of interlayer resistance experimentally, and compared them with our previous model [13].

The layered organic conductor *α*-(BEDT-TTF)$_2$I$_3$, has attracted a great deal of attention as a 2D massless Dirac fermion system following graphene [14]. Because the coupling between BEDT-TTF conducting layers is very small (interlayer transfer energy $t_c$ is much less than $1\,\text{meV}$), this compound is usually regarded as a 2D system. At ambient pressure, it undergoes a metal-insulator transition into a charge-ordered phase at $T = 135\,\text{K}$. Under high pressure ($P > 1.5\,\text{GPa}$), this transition is suppressed, so that the metallic phase survives down to low temperatures [15]. According to the tight-binding calculation in the metallic phase, each BEDT-TTF layer has 2D band dispersion in which the conduction and valence bands contact at two points forming a pair of Dirac cones as shown in the inset of Fig. 1(b) [16,17]. In contrast to graphene, a pair of tilted and anisotropic Dirac cones exists at two general points $\mathbf{k}_0$ and $-\mathbf{k}_0$ in the 2D Brillouin zone, forming valleys. The Fermi level is fixed at the Dirac point because of crystal stoichiometry. In the field above $0.2\,\text{T}$, we can reach the quantum limit, where the Fermi level is located only in the $n = 0$ LL, as the cyclotron energy becomes large around the Dirac point.

α-(BEDT-TTF)$_2$I$_3$ shows anomalous transport behaviors in the quantum limit; the negative interlayer magnetoresistance (MR) [18] and the anomalous interlayer Hall effect [19]. These phenomena have been well explained as magnetotransport of the multilayer Dirac fermion system in the quantum limit, assuming the perturbative interlayer coupling [20] and no degeneracy breaking of the $n = 0$ LL [21,22]. The Dirac cones and van Hove singularity in α-(BEDT-TTF)$_2$I$_3$ have been investigated indirectly using magnetotransport [23-25], specific heat [26], thermopower [27], and NMR [28] measurements.

In stronger magnetic fields, the interlayer MR $R_{zz}$ shows remarkable behaviors [29]. After showing negative MR, it turns to positive with a local minimum. This suggests the splitting of the $n = 0$ LL due to the degeneracy breaking, which decreases the density of states at the Fermi level [21]. After showing the minimum, $R_{zz}$ increases exponentially obeying $R_{zz} \propto 1/\exp(-\mu_B B/k_B T)$ at $T > 1$ K [29]. This activated transport suggests that the Fermi level is located in a mobility gap between the split levels of the $n = 0$ LLs. These data can only indicate the $\nu = 0$ QH state [30]. In stronger fields, the exponential increase of $R_{zz}$ tends to saturate to a more moderate curve [29]. The saturation after activated behavior indicates the existence of a weak transport channel other than the insulating bulk channel. It strongly suggests the existence of edge channels originating from the QHF state [30].

Based on this scenario, we theoretically demonstrated the characteristic interlayer edge transport in multilayer QHF [13]. In a multilayer system such as $\alpha$-(BEDT-TTF)$_2$I$_3$, the helical edge states in the 2D layers (Fig. 1(b)) couple with each other, forming the helical surface state surrounding the crystal as shown in Fig. 1(c). The helical surface state could contribute to the metallic interlayer transport. In QHF, the bulk region inside the crystal shows the insulating activated transport, because the Fermi level is located in the bulk mobility gap. Therefore, the surface transport dominates the interlayer transport, limiting the exponential increase of the interlayer MR. In addition, it has remarkable dependence on the magnetic field orientation, reflecting the resonant tunneling between the helical edge states on neighboring layers. When the magnetic flux penetrates between the helical edge channels on neighboring layers, the interlayer electron tunneling shifts the center coordinates $x_0$ as shown in Fig. 1(a), so that the tunneling process generally does not conserve energy. Therefore, the tunneling between the helical edge states on neighboring layers becomes possible only when the magnetic field is parallel to one of the side surfaces of the crystal. According to this model [13], the lowest order contribution of the interlayer transfer $t_c$ to the interlayer surface conductivity $\sigma_{zz}^{(\text{surface})}$ is given by

$$\sigma_{zz}^{(\text{surface})} = \frac{4t_c^2 c \tau^{(\text{edge})}}{\hbar^2 v_F^{(\text{edge})}} \left(\frac{e^2}{h}\right) \frac{1}{1+(B_x/B_0)^2}. \tag{1}$$

Here, we considered only one side surface parallel to the $yz$-plane, and $c$, $\tau^{(\text{edge})}$, and $v_F^{(\text{edge})}$ denote the interlayer spacing, the scattering time of the edge channel, and the group velocity of the edge channel, respectively. The characteristic field $B_0$ is defined by $B_0 \equiv h/(2\pi e c v_F^{(\text{edge})} \tau^{(\text{edge})})$. We note that the value of $\sigma_{zz}^{(\text{surface})}$ is much less than $e^2/h$ at the limit of $T=0$, because $v_F^{(\text{edge})} \ll v_F$, $2t_c c/\hbar \ll v_F$, and $t_c \ll \hbar/\tau^{(\text{edge})}$, where $v_F$ is the group velocity of the Dirac fermions in each layer. This is also a remarkable feature of the helical surface state as a 2D electron system. Above features are expected from the analogy with the chiral surface state in the $\nu \neq 0$ QH multilayer systems [31-33].

In actual crystals, the side surfaces face various directions. So, the total interlayer surface conductance must take the maximum value at the field direction around the common axis parallel to all side surfaces. The total interlayer resistance is given by

$$R_{zz} = 1 \bigg/ \left(\sigma_{zz}^{(\text{bulk})} \frac{S}{L_z} + \sum \sigma_{zz}^{(\text{surface})} \frac{L^{(\text{edge})}}{L_z}\right), \tag{2}$$

where $S$, $L^{(\text{edge})}$, and $L_z$ are the cross-sectional area, the length of each edge, and the thickness of a slab-shaped crystal, respectively. $\sigma_{zz}^{(\text{bulk})}$ is the bulk interlayer conductivity given in Eqs. (12) in Ref. 21. The summation is taken for all side surfaces surrounding the crystal. As the field and temperature dependence of $\sigma_{zz}^{(\text{surface})}$ is weak, $\sigma_{zz}^{(\text{surface})}$ becomes dominant in the denominator of $R_{zz}$ at low temperatures or for high fields where $\sigma_{zz}^{(bulk)}$ is exponentially small, resulting in the saturation of $R_{zz}$.

In the following section, we present the experimental evidence for the surface transport in $\alpha$-(BEDT-TTF)$_2$I$_3$, and confirm that it originates from the helical surface state. Single crystals of $\alpha$-(BEDT-TTF)$_2$I$_3$ were grown by the conventional electrochemical method. For the interlayer transport experiments, the electrodes are formed on the top and bottom surfaces of the slab-shaped crystal by gold evaporation. Samples were mounted in the piston-cylinder-type pressure cell set to a rotating probe in a superconducting magnet. To suppress the charge ordering and achieve the Dirac state, we applied $1.7-1.8$ GPa of pressure at room temperature. The pressure was monitored using the resistance of a manganin wire. $R_{zz}$ was measured with the quasi-four-terminal method using a DC current parallel to the stacking $\mathbf{c}$-axis ($z$-axis). Since the electrodes cover the top and bottom surfaces, the effect of the current jetting is ruled out [34]. The current value was changed within $10$ nA $-10$ $\mu$A range depending on the resistance, so as to ensure that it stays in the Ohmic region, and the effect of nonlinear transport is also ruled out.

First, we show that surface transport exists in $\alpha$-(BEDT-TTF)$_2$I$_3$ when the interlayer MR shows saturating behavior. We examined whether the interlayer MR scales with the sample cross-sectional area or not. For this purpose, we cut one slab-shaped crystal into two pieces with different cross-sectional areas and perimeters but the same thickness, and compared their interlayer MR. Figures 2(a) and 2(b) exhibit the temperature and magnetic field dependence of interlayer resistance normalized by the cross-sectional area, $R_{zz}S$, for Samples #1 (solid curves) and #2 (dashed curves) at $P = 1.7$ GPa. The cross-sectional areas and perimeters ($L = \sum L^{(\text{edge})}$) of #1 and #2 were determined from the microscope images as $S_1 = 0.219$ mm$^2$, $L_1 = 1.89$ mm, $S_2 = 0.0675$ mm$^2$, and $L_2 = 1.13$ mm. The common thickness was $L_z = 0.05$ mm. The magnetic field was parallel to the normal of conducting layers (z-axis).

In Fig. 2(a), the interlayer resistance shows a metallic temperature dependence at high temperatures; as the temperature decreases, it first increases exponentially, and then tends to saturate at high magnetic fields. Above the saturation temperature, the values of normalized resistance $R_{zz}S$ of the two samples (solid and dashed curves) almost coincide with each other. This means that the interlayer resistance is scaled by the cross-sectional area as $R_{zz} = \left(\sigma_{zz}^{(\text{bulk})} S / L_z\right)^{-1}$, indicating uniform bulk conduction. Note that $R_{zz}S$ is plotted on a logarithmic scale in the figure. Although we can see a small mismatch between the two samples in the low resistance region, it might originate from the contact resistance due to the quasi-four-terminal measurement. In contrast, $R_{zz}S$ of the two samples clearly takes different values in the saturation region. This indicates that the saturation originates from the nonuniform local transport, and the appearance of surface transport is a plausible explanation. If this is the case, $R_{zz}$ must be scaled by the perimeter $L$. However, $R_{zz}$ does not necessarily show this clear scaling relation, because the side surface is not flat but too jagged to estimate the precise perimeter.

The field dependence in Fig. 2(b) also shows similar behavior. With increasing magnetic fields, $R_{zz}$ first decreases (negative MR), then increases exponentially, and finally shows saturation. Although the $R_{zz}S$ curves of the two samples almost overlap at low magnetic fields, they deviate from each other in the saturation region. This indicates that the saturation in the high field region does not originate from uniform bulk transport, and strongly suggests the appearance of surface transport.

Supposing that the saturation originates from surface transport, the surface conductivity $\sigma_{zz}^{(\mathrm{surface})} = L_z / R_{zz} L$ is estimated as $0.04\, e^2/h$ for Sample #1 and $0.07\, e^2/h$ for #2 at $B = 13\,\mathrm{T}$ and $T = 1.0\,\mathrm{K}$. Since these values are much smaller than $e^2/h$, it is consistent with the transport of the helical surface state previously mentioned.

Next, we investigated the dependence of interlayer MR on the magnetic field orientation using the rotating probe with the pressure cell. The area and perimeter of the plate-like sample was $S = 0.11\,\mathrm{mm}^2$ and $L = 2.00\,\mathrm{mm}$, respectively, and the thickness was $L_z = 0.05\,\mathrm{mm}$. Figure 3(a) illustrates the measured interlayer resistance $R_{zz}$ as a function of magnetic field orientation and strength at $T = 1.5\,\mathrm{K}$ and $P = 1.8\,\mathrm{GPa}$. The distance and orientation from the origin indicate the strength and orientation of the magnetic field, respectively. The color indicates the value of $R_{zz}$. The $B_z$-axis is taken to be normal to the 2D conducting layers, and the $B_\parallel$-axis corresponds to an in-plane direction chosen arbitrarily. In general field orientations, $R_{zz}$ increases with the field strength, as shown by the color change from blue to red. Around the polar angle $\theta \equiv \tan^{-1}(B_\parallel / B_z) = 15°$, however, we can see a radial white belt along the dashed line in the figure, which indicates the saturation of $R_{zz}$ to the white value. This result shows that the saturation of $R_{zz}$ occurs at a specific field orientation. This specific orientation is different in every experiment using different samples, suggesting that it is not characteristic to the material but depends on the configuration of the sample's side surfaces.

According to the model of the multilayer QHF [13], the interlayer surface transport shows a resonant increase when the field is parallel to most of the side surfaces. This causes the saturation of $R_{zz}$ at a specific field orientation. Figure 3(b) shows the simulated angle-dependence of $R_{zz}$ at $T=0$ [13], where it is assumed that the side surface normal to the $x$-axis dominantly contributes to $R_{zz}$. Here, the interlayer resistance $R_{zz}$ is normalized by $R_0 \equiv (\hbar^2 cL_z / 4t_c^2 \tau^{(\text{bulk})2} S)/(e^2/h)$, and $c=1.75$ nm, $v_F = 2.4 \times 10^4$ m/s, $\tau^{(\text{bulk})} = 2$ ps, $\tau^{(\text{edge})} = 20$ ps, and $cL^{(\text{edge})}/S = 10^{-4}$ are used following Ref. 13. We can see a white belt along the $B_z$-axis. It reflects the resonant increase of $\sigma_{zz}^{(\text{surface})}$ when $B_x = 0$, since the finite $B_x$ suppresses the tunneling between helical edge states. The observed angle-dependent feature in Fig. 3(a) is consistent with this model.

In summary, we have presented experimental results that strongly suggest the realization of the $v=0$ multilayer QHF state accompanied by the helical surface state in the organic Dirac fermion system α-(BEDT-TTF)$_2$I$_3$. The interlayer MR shows saturating behavior after an exponential increase due to activated transport in high magnetic fields and low temperatures. These features suggest the $v=0$ QH state with metallic edge channel. We found that the saturating interlayer MR is not scaled by the sample cross-sectional area, indicating nonuniform transport in the saturation region. The surface transport due to the helical surface state in the QHF is one of the plausible mechanisms. Moreover, we found that the saturating resistance shows resonant decrease in the field orientation parallel to most of the side surfaces of the sample. This feature is well explained by the surface transport due to the helical surface state in the QHF. Therefore, the realization of the $\nu=0$ QHF state with the helical surface state is strongly suggested in $\alpha$-(BEDT-TTF)$_2$I$_3$. This is the first observation of the topological phase in organic molecular crystals, although topological phases have been theoretically proposed [35,36].

The authors thank Prof. N. Tajima, Prof. K. Kajita, Prof. A. Kobayashi, and Prof. K. Kanoda for valuable discussions and comments. This work was supported by JSPS KAKENHI Grant Numbers JP23110705, JP25107003, JP15K21722, and JP16H03999.

*corresponding author, osada@issp.u-tokyo.ac.jp

#present address: *International Center for Material Nanoarchitectonics, National Institute for Materials Science, 1-1 Namiki, Tsukuba, Ibaraki 305-0044, Japan*.

**83**, 072002 (2014).

[15] N. Tajima, S. Sugawara, M. Tamura, Y. Nishio, and K. Kajita, J. Phys. Soc. Jpn. **75**, 051010 (2006).

[16] S. Katayama, A. Kobayashi, and Y. Suzumura, J. Phys. Soc. Jpn. **75**, 054705 (2006).

[17] A. Kobayashi, S. Katayama, Y. Suzumura, and H. Fukuyama, J. Phys. Soc. Jpn. **76**, 034711 (2007).

[18] N. Tajima, S. Sugawara, R. Kato, Y. Nishio, and K. Kajita, Phys. Rev. Lett. **102**, 176403 (2009).

[19] M. Sato, K. Miura, S. Endo, S. Sugawara, N. Tajima, K. Murata, Y. Nishio, and K. Kajita, J. Phys. Soc. Jpn. **80**, 023706 (2011).

[20] T. Osada and E. Ohmichi, J. Phys. Soc. Jpn. **75**, 051006 (2006), and references therein.

[21] T. Osada, J. Phys. Soc. Jpn. **77**, 084711 (2008).

[22] T. Osada, J. Phys. Soc. Jpn. **80**, 033708 (2011).

[23] S. Sugawara, M. Tamura, N. Tajima, R. Kato, M. Sato, Y. Nishio, and K. Kajita, J. Phys. Soc. Jpn. **79**, 113704 (2010).

[24] M. Monteverde, M. O. Goerbig, P. Auban-Senzier, F. Navarin, H. Henck, C. R. Pasquier, C. Meziere, and P. Batail, Phys. Rev. B **87**, 245110 (2013).

[25] A. Mori, M. Sato, T. Yajima, T. Konoike, K. Uchida, and T. Osada, Phys. Rev. B **99**, 035106 (2019).

[26] T. Konoike, K. Uchida, and T. Osada, J. Phys. Soc. Jpn. **81**, 043601 (2012).

[27] T. Konoike, M. Sato, K. Uchida, and T. Osada, J. Phys. Soc. Jpn. **82**, 073601 (2013).

[28] M. Hirata, K. Ishikawa, K. Miyagawa, M. Tamura, C. Berthier, D. Basko, A.

**Figure 1** (Sato *et al.*)

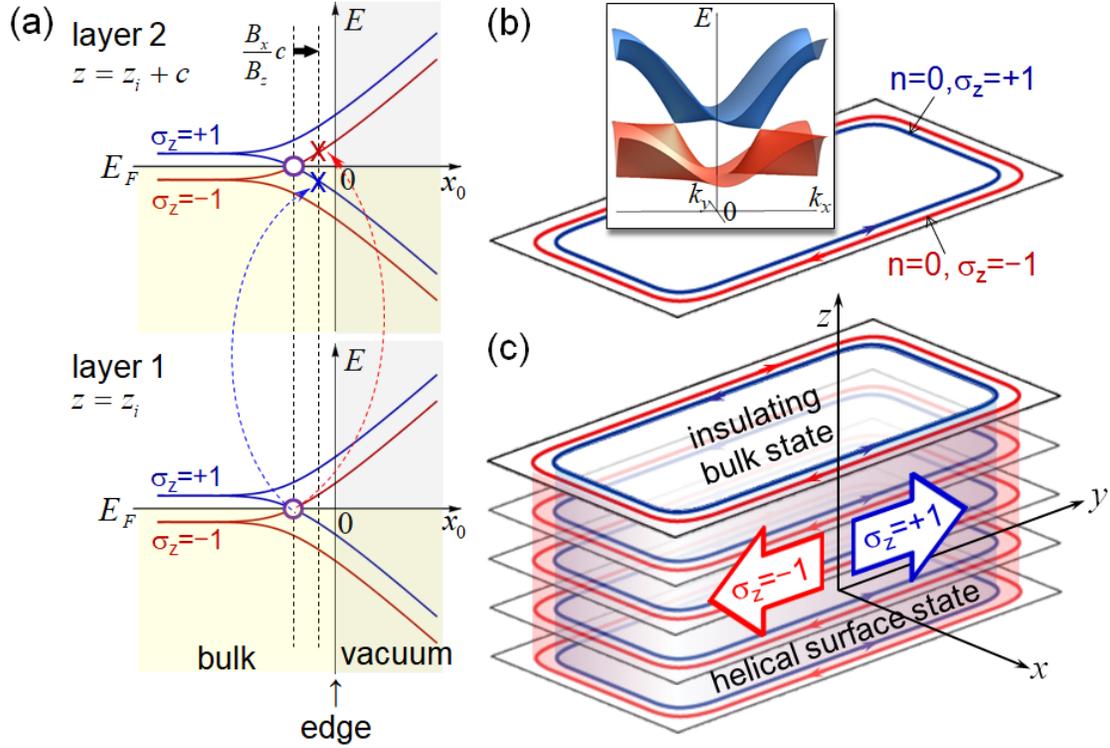

**FIG. 1.** (color online)

(a) Helical edge states on two neighboring QHF layers. The open circle indicates a helical edge channel. The dashed arrows show allowed interlayer tunneling between the edge states under the tilted magnetic field with finite $B_x$, which is the normal component to the surface. (b) Schematic of the helical edge channel in the 2D QHF. Inset shows the band dispersion in the conducting layer of $\alpha$-(BEDT-TTF)$_2$I$_3$. (c) Schematic of the helical surface state in the multilayer QHF.

**Figure 2** (Sato *et al.*)

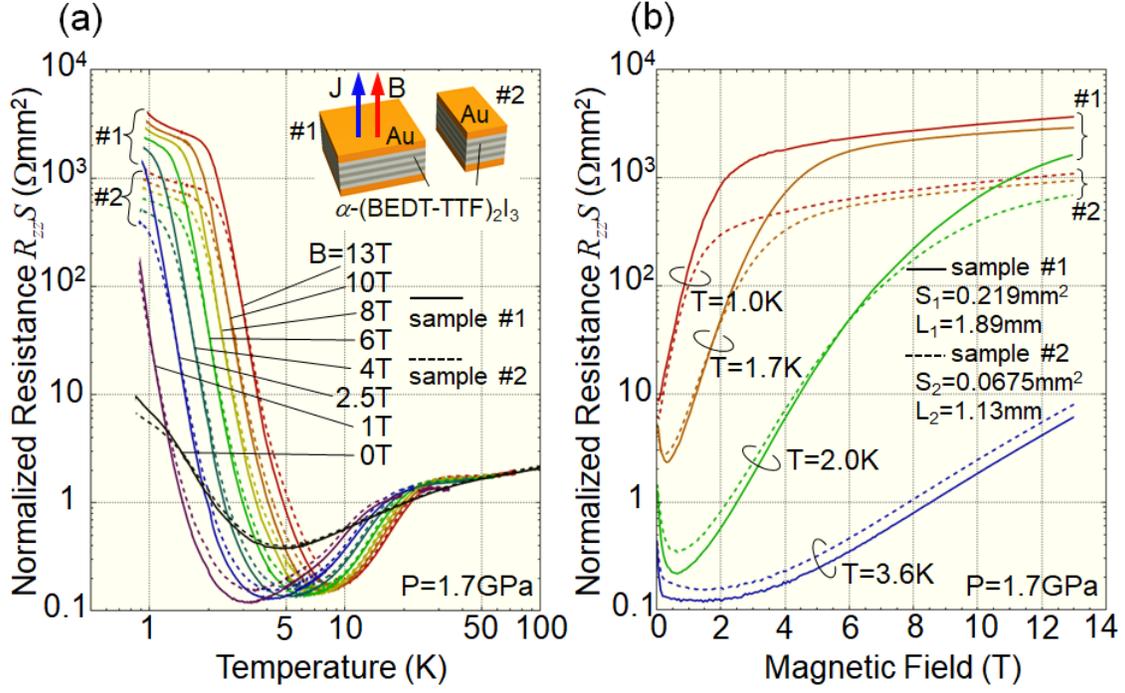

**FIG. 2.** (color online)

Interlayer resistance $R_{zz}$ normalized by cross-sectional area $S$ of two $\alpha$-(BEDT-TTF)$_2$I$_3$ crystals under $P = 1.7$ GPa. The solid and dashed curves represent the two samples (#1 and #2) with different $S$. (a) Temperature dependence of $R_{zz}S$ at fixed magnetic fields. The inset illustrates the experimental configuration. (b) Magnetic field dependence at fixed temperatures.

**Figure 3** (Sato *et al.*)

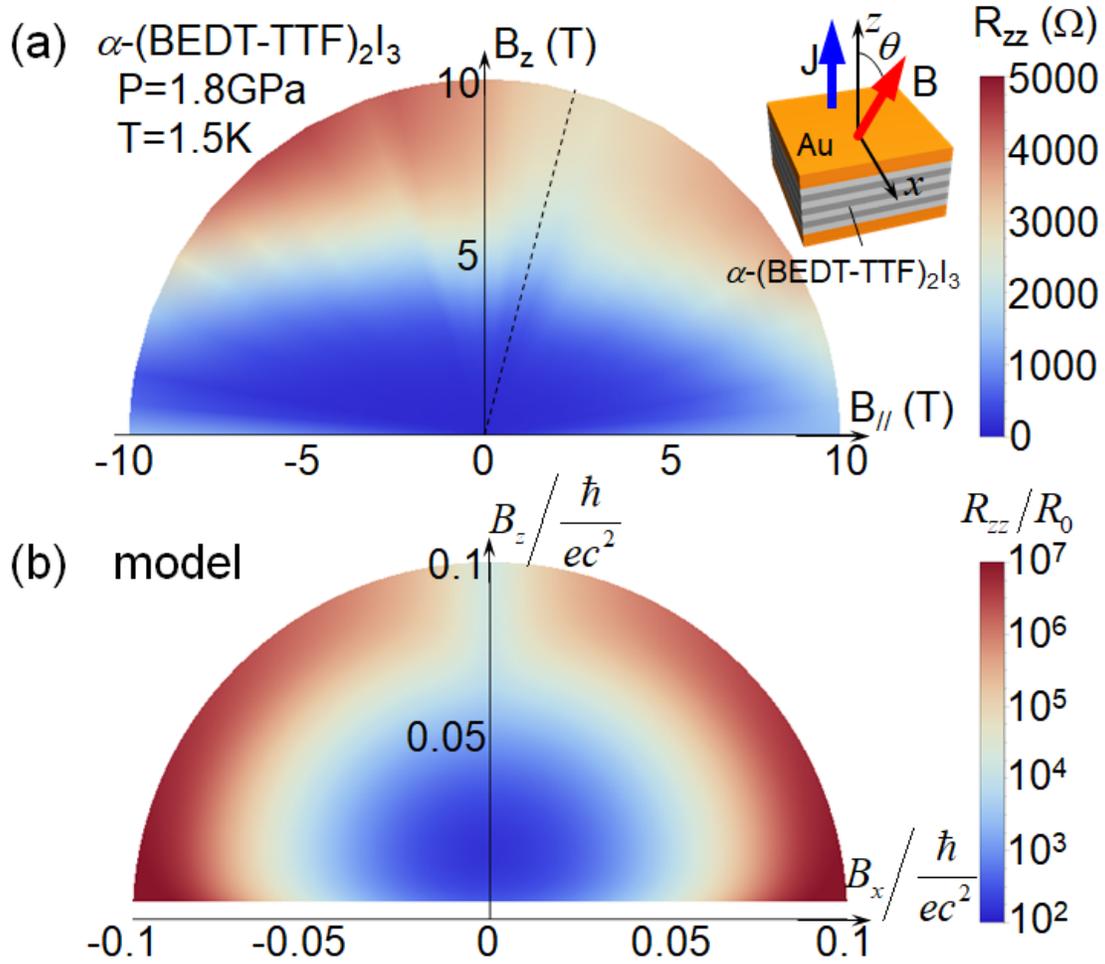

FIG. 3. (color online)

(a) Dependence of the interlayer resistance of α-(BEDT-TTF)$_2$I$_3$ on the strength and orientation of the magnetic field at $P = 1.8$ GPa and $T = 1.5$ K. The color indicates the interlayer resistance. The inset shows the experimental configuration. (b) Simulation of the interlayer resistance at $T = 0$ based on the multilayer QHF model following Ref. 13.

17